\documentclass[a4paper]{jpconf}
\usepackage{graphicx}
\begin{document}
\title{Radiowave neutrino detection \\
{\normalsize (ARENA06 Conference Summary talk,
 Newcastle, UK, June 28--July 1, 2006)}}
%\title{Preparing a paper using \LaTeXe\ for publication in \jpconf}

%Title of paper
%\begin{verbatim}
%\title{Status of radiowave detection of neutrinos \\ (ARENA06 Conference Summary talk, Newcastle, UK, June 28--July 1, 2006)}
%\end{verbatim}

\author{Dave Z. Besson}
\address{U. Kansas Dept. of Physics and Astronomy, Lawrence KS  66045-2151}
%\maketitle

%\begin{abstract}The current (Sept. 30, 2006) status of radiowave detection of electromagnetic showers is summarized. I present my personal view of the relative strengths of the various approaches being pursued. \end{abstract}

%\normalsize\baselineskip=15pt

\begin{quote}
{\it ``I am often asked how radio works. Well, you see, wire telegraphy is
like a very long cat. You yank his tail in New York and he meows in Los
Angeles. Now, radio is exactly the same, except
that there is no cat.'' (Attributed to A. Einstein)}
\end{quote}

\section{Introduction}
Depending on whether you search www.google.com, www.google.it, or
www.google.ru, radiofrequency (RF) signal
transmission was
first developed at the turn of the century by either Gugliemo Marconi
in June, 1896,
(see the Wikipedia pages in Italian),
or Alexander Popov (see the Wikipedia pages
in Russian) in May, 1895. Nikola
Tesla, working about 400 km due
east of the University of Kansas along
Interstate 70, might rightfully be considered the progenitor of the
modern-day wireless industry; his innovative suggestions regarding
the possible propagation of
surface waves
has also been the subject of neutrino
detection schemes\cite{jpr}, among others\cite{TheWhiteStripes}.

Since radio wavelengths are `macroscopic', for several applications
they have advantageous signal production and transmission properties. 
For a typical antenna,
the effective height, which is directly proportional to the
voltage produced by an incident electric field, is of order the antenna
length. Machining a piece of copper of scale 1 m is considerably
easier than machining a piece of copper two orders of magnitude larger
or smaller. 
At meter-scale radio wavelengths (in comparison to
atomic dimensions) dielectric media with only atomic-scale impurities
may be
largely radio-transparent.
For example, the attenuation length of radiowaves in 
South Polar ice has been measured to be $\sim$1.5--2 km over
the frequency range 200 MHz - 800 MHz at the South Pole\cite{iceatten}.

For these and other reasons, the initial assertion by Askaryan\cite{Askaryan61}
that high-energy electromagnetic showers will produce RF signals
in dense media stronger than their optical counterparts, 
has (inevitably) engendered
a very recent flurry of activity in this field. Whereas, 
a decade ago, there was only one active experiment (RICE) seeking
detection of ultra-high energy (UHE) showers in dense media
via radio detection,
there are now several active groups. 

\section{Motivation}
The GZK-effect\cite{GZK0} implies that 100 EeV
charged particles arriving at Earth
must originate from within a sphere
of radius $\sim$10 Mpc. This is a remarkably
tiny volume, corresponding roughly to the scale of the Local
Galactic Group. In terms of redshift, relative to the 14 Gyr history
of the Universe, we are sensitive only to those Ultra-High
Energy Cosmic Rays (UHECR) produced in the last 30 Myr. If the
Universe evolved through a past epoch where particle acceleration
mechanisms were present in the past that have not been active during
last 0.3\% of the temporal history of the Universe, we could not
observe the resulting UHECR's. Neutrinos suffer no GZK-attenuation and
therefore, in principle, probe all redshifts.

Detection of ultra-high energy neutrinos offers
unique insights into the most energetic processes
in the Universe.
Super-massive black holes (Active Galactic Nuclei), 
the collapse of bizarre twists of spacetime left 
over from the Big Bang (topological defects) and 
Weakly Interacting Massive Particles (WIMP's) are 
only some of the sources that would be probed by 
an ultra-high energy electromagnetic (and 
hadronic) shower detector. There is 
also a great deal of particle physics information 
which can be deduced, for instance, from the 
angular distribution of upward-coming neutrino 
events through different chords of the earth. 
As recently emphasized by proponents of ARIANNA\cite{ARIANNA},
such data 
could be used to measure
weak cross-sections at energies unreachable by 
any man-made accelerator, and thereby probe the 
internal structure of protons at length scales 
orders of magnitude smaller than those currently achievable.
This capability 
of probing such minute length scales may also 
reveal new physics (extra dimensions, micro-black-holes)\cite{McKayBH} 
for the first time.

\section{Radio signals from neutrino-generated showers in dielectrics}
%Askaryan\cite{Askaryan} subsequently predicted a net charge imbalance in dense media and coherent radio power scaling like the energy of the shower squared, providing impetus for continued progress in UHE air shower studies\cite{LOTHAR,Rosner}. Currently, there are ongoing searches for $E>10^{20}$ eV cosmic rays and neutrinos impinging on the Moon (GLUE\cite{Moon}), with future plans for salt-based neutrino detection (SALSA\cite{SALSA}). The coherent enhancement of Cherenkov radiation in the microwave region has been observed in the laboratory\cite{Takahashi, Wake}. The Argonne wake-field acceleration project\cite{Wake} has successfully generated extremely large microwave field-strengths by manipulating coherent radiation from an intense electron beam. Most recently, a test beam experiment at SLAC by Gorham and Saltzberg has given the most direct confirmation of the Askaryan effect\cite{SLAC-beam-test}.
For definiteness, consider a $\nu_e$ undergoing a 
charged current interaction, $\nu_e+N\to e+N'$. The primary UHE 
neutrino transfers most of its 
energy to the electron, which quickly builds an 
exponentially increasing shower of $e^+e^-$ pairs.  The number of pairs 
$N_e$ scales like the primary energy.  In the most populated region of the 
shower, at the ``bottom" of its energy range, a charge imbalance 
develops as positrons drop out and atomic electrons scatter in.  
Detailed Monte Carlo calculations by 
Zas, Halzen and Stanev (ZHS)\cite{ZHS} 
and supported by subsequent GEANT simulations\cite{Soeb} and
subsequent refinements of ZHS\cite{Jaime-papers},
find that the net charge of the 
shower is about 20\% of $N_e$.  The electric field produced by this 
relativistic pancake is dominated by coherent Cherenkov radiation for
wavelengths in the radio frequency region.  
Equivalently, for wavelengths large comparable to the transverse size of the
shower ($\sim 2r_{Moliere}$, or $\sim$10cm in ice),
the relativistic
pancake can be treated as a single, extended, radiating charge.
(Clearly, in the limit $\lambda\to\infty$, the radiating region approaches a
point charge.) 

%The combination of this coherent magnification of the signal strength, plus the RF clarity of cold ice, has prompted several new experimental initiatives, based on radio-wave detection of neutrino-induced electromagnetic cascades. These initiatives seek to extend the high-energy neutrino astronomy frontier beyond the experimental reach of the photomultiplier-based detectors.

\subsection{Laboratory Verification}
\begin{figure}[h]
\begin{minipage}{18pc}
\includegraphics[width=16pc,angle=0]{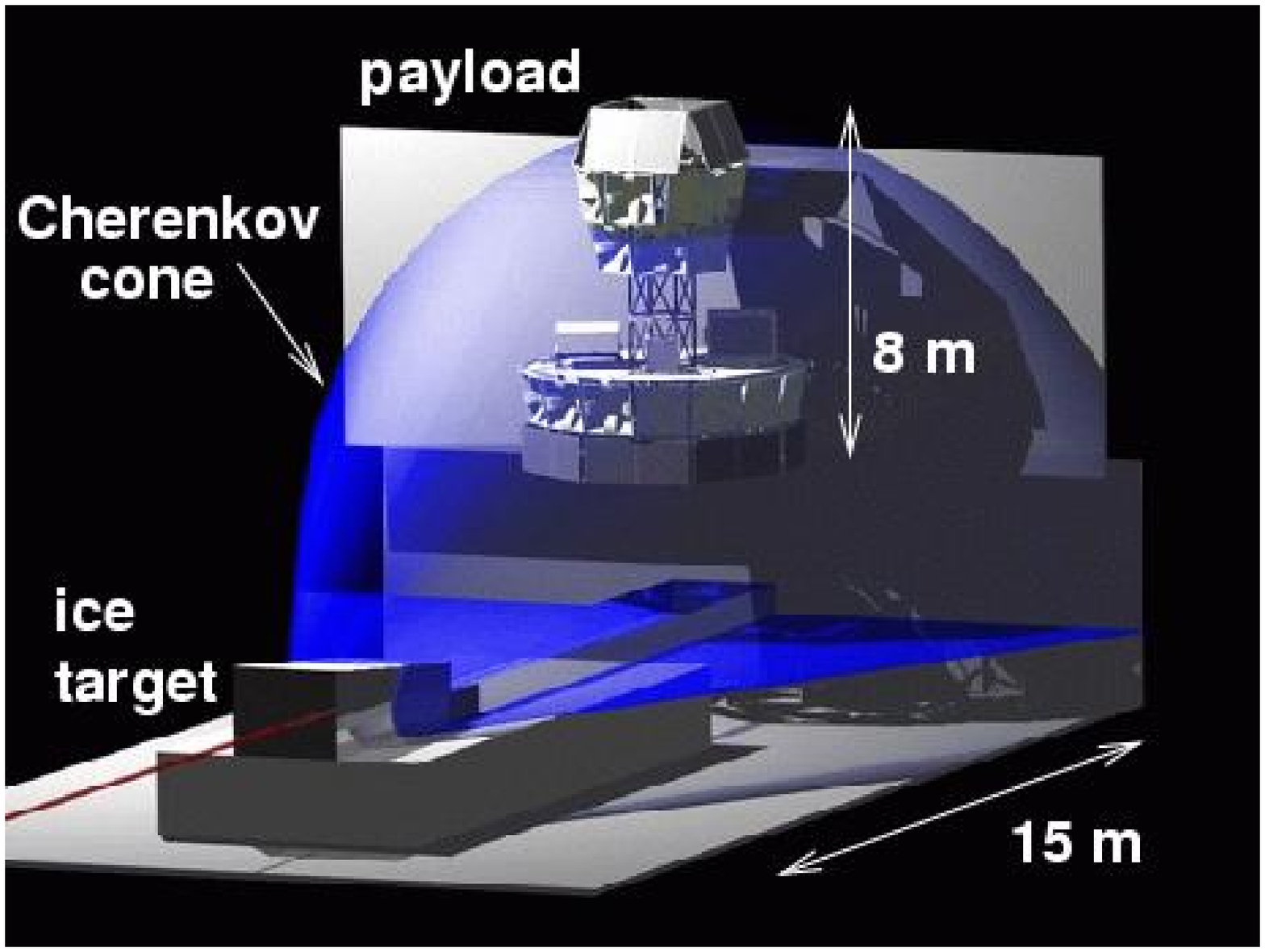}
\caption{\label{fig:ANITAschematic}ANITA SLAC testbeam geometry.}
\end{minipage}
\hspace{0pc}
\begin{minipage}{20pc}
\includegraphics[width=18pc,angle=0]{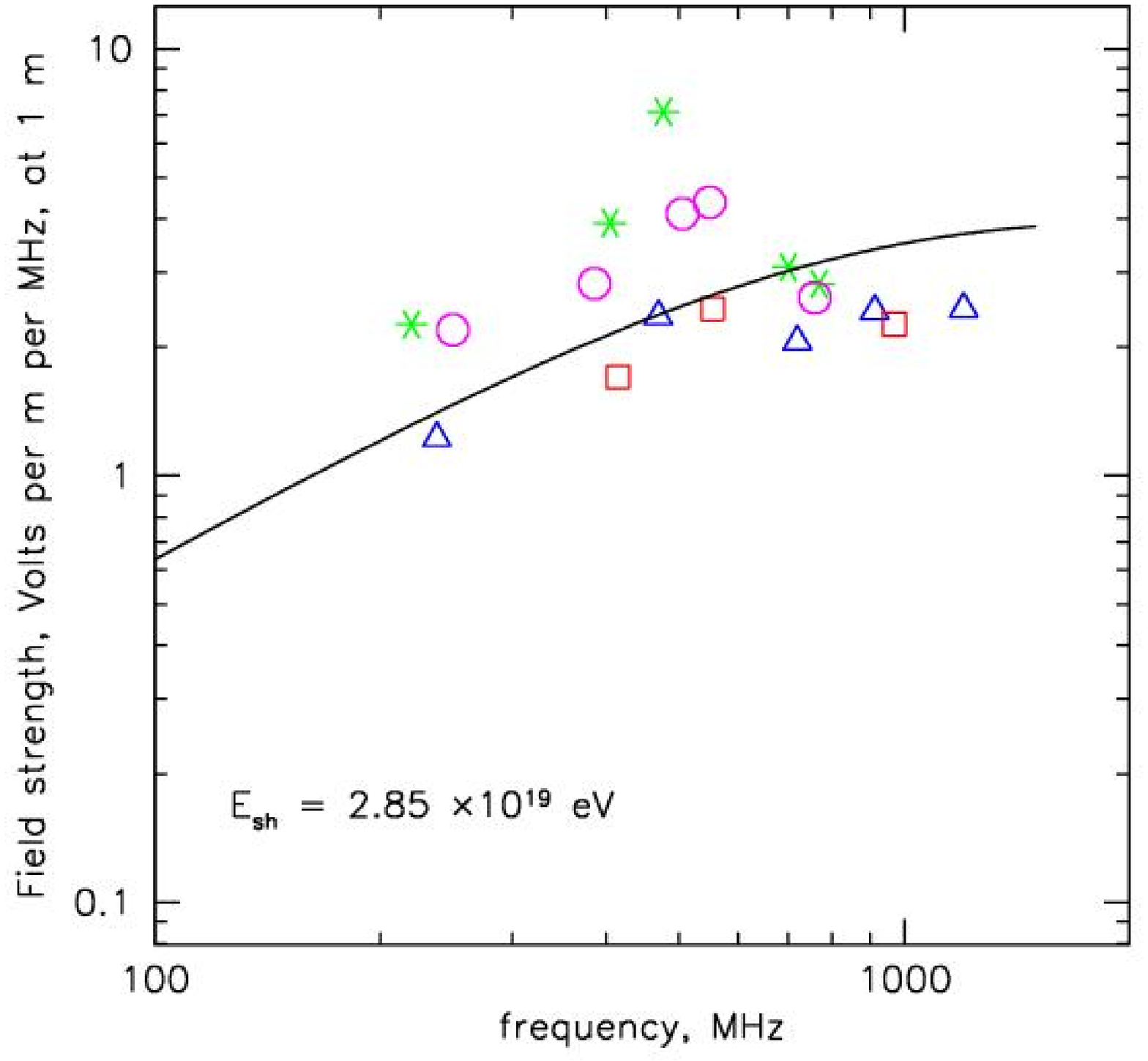}
\caption{\label{fig:SigVfreq}Measured voltage vs. frequency.}\end{minipage} 
\end{figure}

Laboratory tests of the radio signal resulting from a charged particle beam
have been ongoing for the last decade, with the most definitive tests
occurring in the last five years. The GLUE/ANITA team has performed
a sequence of measurements, beginning with an electron beam incident
on a sand target\cite{testbeam1}, a photon beam on a salt target\cite{testbeam2}, and most recently, 
ANITA has calibrated their antenna response in a
SLAC testbeam on an ice target\cite{testbeam3}, 
as shown in Figure \ref{fig:ANITAschematic}. The signals
observed in a variety of radio antennas and receivers track the
expected signal strength (Fig. \ref{fig:SigVfreq}), near-field
complications notwithstanding. Surface roughness effects
are relevant to both the beamtest as well as the flight itself.
Studies of signal transmission through
the firn, as well as data scaled-up from a clever optical
wavelength set-up, indicate that surface roughness should, if anything,
enhance the number of detected events. 

\subsection{Active Experiments}
Initial projects exploring radio Cherenkov signals at the South Pole (RAND\cite{RAND})
or at Vostok\cite{Alexey} date back to the early-90's.
Currently, two Antarctic projects (ANITA\cite{ANITAprl} and RICE\cite{astroph06})
are in the data-taking phase. 
ANITA, largely funded by NASA as part of the LDB (Long Duration
Balloon) program, will launch from McMurdo Base, Antarctica,
sometime around December 1, 2006 for an extended circumpolar
flight. 
Two prototype ANITA antennas mounted on the TIGER payload during December 2003
provided essential experience and data, in preparation for the main
flight.
During its planned 45-day December-January flight of this year, comprising
three full revolutions around the Antarctic continent, 
ANITA will synoptically (from an elevation of 38 km) 
view a cylindrical volume of ice 700 km in radius, 2 km thick on
average, and with an average angular aperture of approximately 0.1 radians. 
With an effective
threshold of $10^{19-20}$ eV (set largely by the typical distance to 
the interaction vertex), ANITA will offer the
best sensitivity to date to the GZK-neutrino
parameter space. 

\begin{figure}[h]
\begin{minipage}{40pc}
\includegraphics[width=24pc,angle=-90]{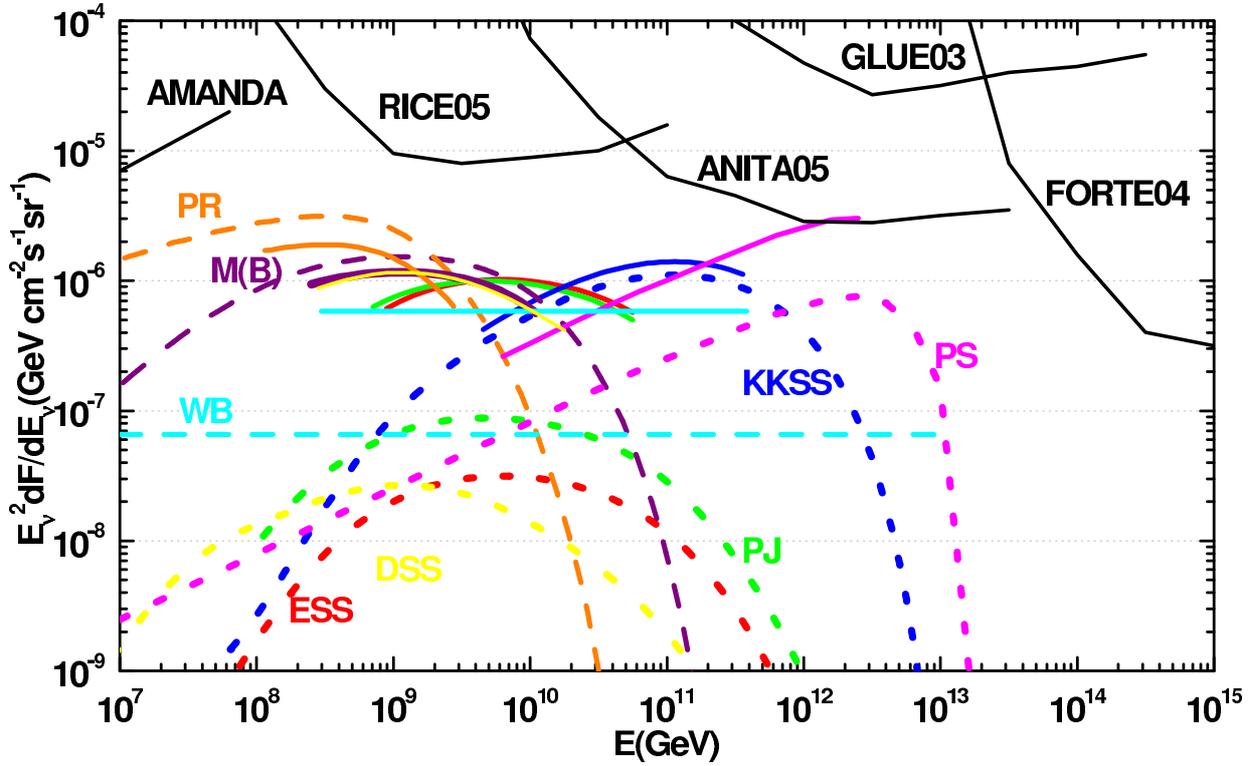}
\caption{\label{fig:limits}Exerimental upper limits on neutrino flux, as a function of energy, for various neutrino flux models.
Upper bounds on total (all flavor) neutrino fluxes for AGN 
models of PR\cite{pr} 
and MB \cite{mb}, GZK\cite{GZK0} 
neutrino models of ESS\cite{ess}, 
PJ\cite{pj}, KKSS\cite{kkss}, and DSS\cite{MSS}, 
and the topological defect model of PS\cite{ps}, due to all flavor 
NC+CC interactions, based on 1999-2005 RICE livetime of about 20500 hrs. 
Dashed curves are for model fluxes and the thick curves are the 
corresponding bounds.  The energy range covered by a bound 
represents the central 80\% of the event rate. Note that
RICE limits are 95\% c.l.; other experimental limits shown are
at 90\% c.l.}
\end{minipage}
\end{figure}

\begin{figure}[h]
%\vspace{-4cm}
\begin{minipage}{20pc}
\includegraphics[width=13pc,angle=-90]{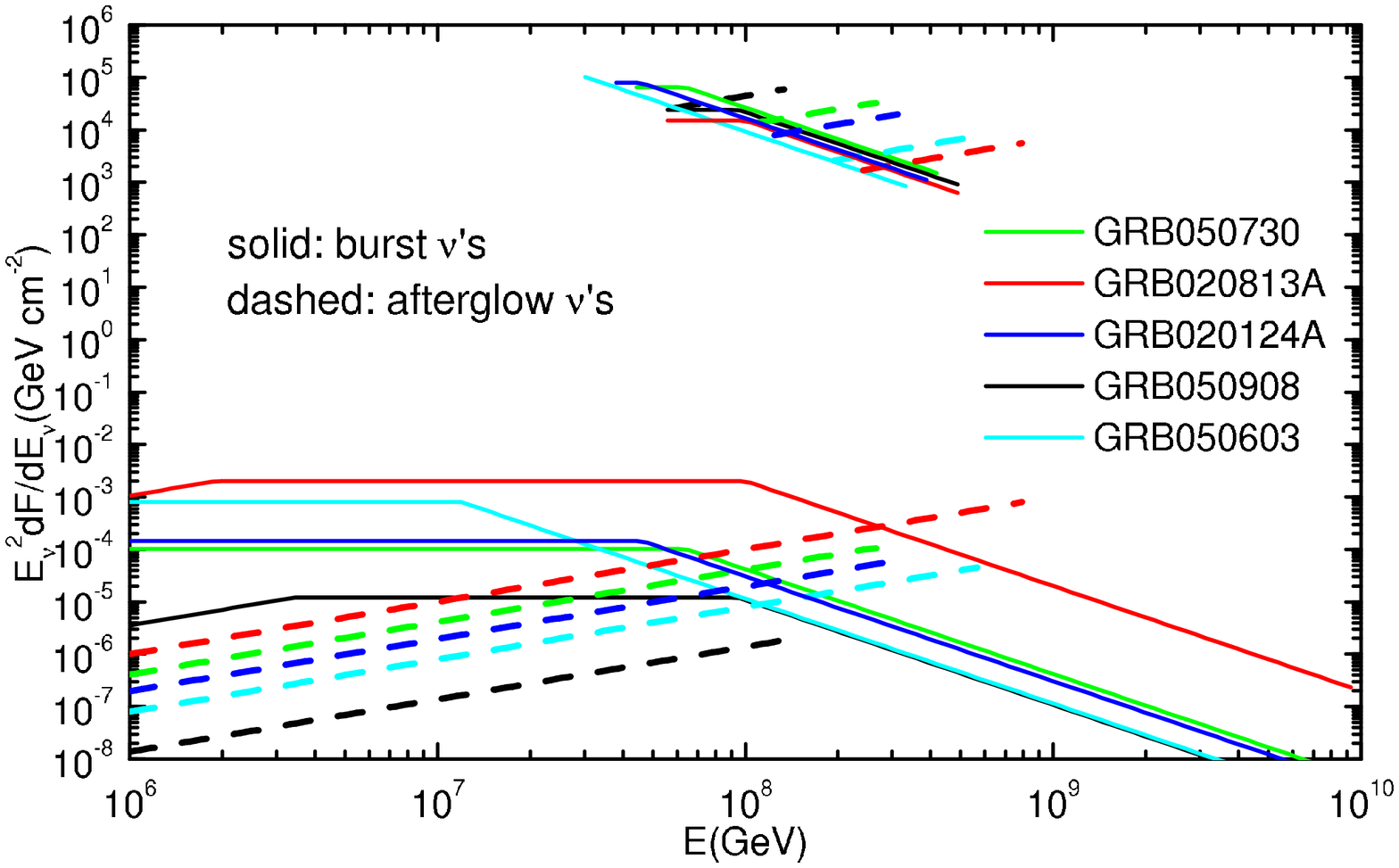}
\caption{\label{fig:ricegrblim}Expected fluxes (solid) and limits (dashed) on UHE neutrinos from GRB's.}
\end{minipage}\hspace{1pc}%
\begin{minipage}{24pc}\includegraphics[width=20pc]{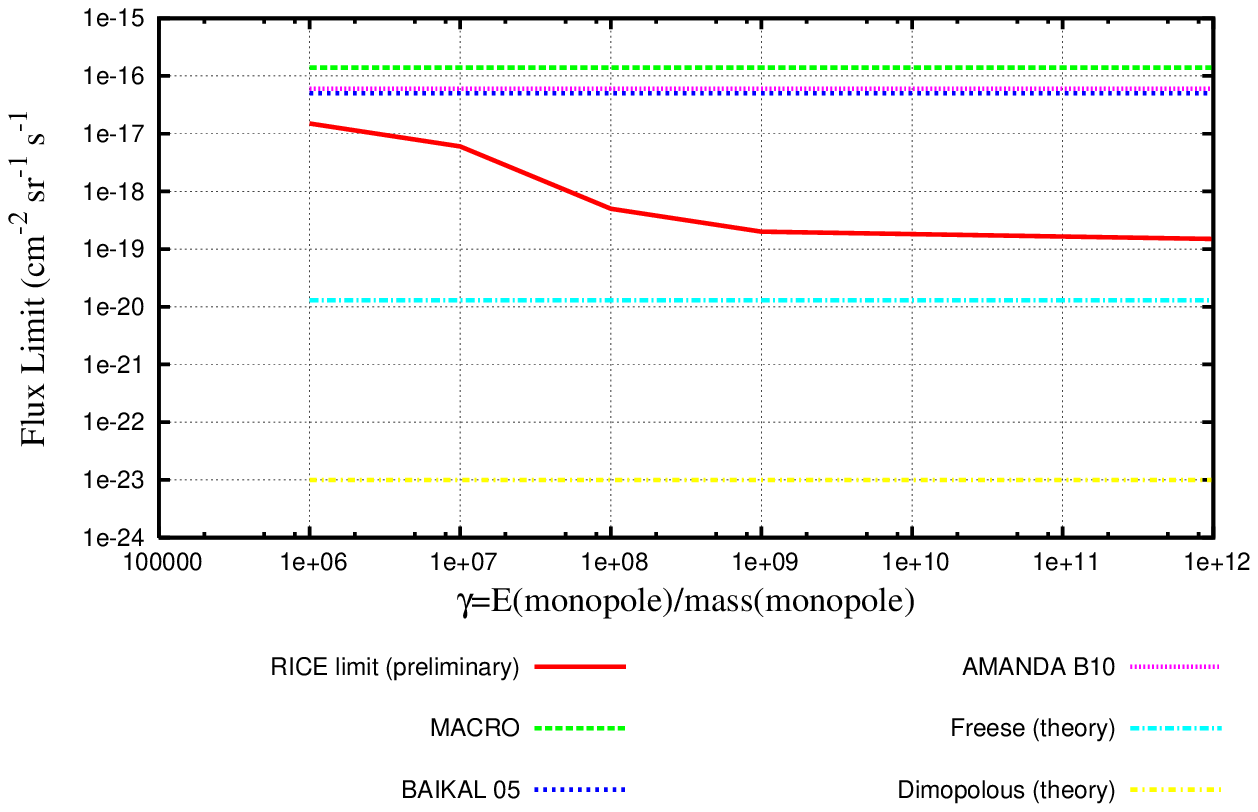}\caption{\label{fig:monopole}RICE monopole flux limits.}\end{minipage} 
\end{figure}

The RICE experiment, which has been taking data since
1999, has a threshold approximately two orders of magnitude lower, with
an effective volume also approximately two orders of magnitude smaller than
ANITA at 100 EeV. Nevertheless, with radio antennas 50-100 meters
apart, RICE will continue to 
have excellent sensitivity in the 100 PeV-100 EeV 
energy regime, not to be superceded until the advent of either the
ARIANNA\cite{ARIANNA} or RICE-successor
(AURA)\cite{AURA}\footnote{The original acronym advocated by
the author for this
experiment (Retrofitted OptiCal SysTem Adapted for Radio) was,
unfortunately, not favored by my collaborators.} experiments. 
A compilation of the current RICE limits on the incident
neutrino flux, superimposed on both experimental
limits as well as theoretical
predictions, is presented in Figure 
\ref{fig:limits}.
The null search for in-ice showers can quickly be translated
into a limit on the diffuse neutrino flux from gamma-ray
bursts\cite{astroph06}. Additionally, a dedicated gamma-ray
burst (GRB)
coincidence study was performed to quantify limits on neutrino-generated
showers from specific GRB's\cite{ricegrb}. 
The GRB sample coincidence sample used was, unfortunately, a
preferentially high-redshift sample. Accodingly,
the limits are rather weak relative to model
expectation (Fig. \ref{fig:ricegrblim}), 
%\begin{figure}[h]\begin{minipage}{30pc}\includegraphics[width=38pc,angle=0]{volexpo.png}\vspace{-7cm}\caption{\label{label}Figure caption for first of two sided figures.}\end{minipage}\end{figure}

As originally pointed out by Wick {\it et al.}\cite{stuart}, a radio detector
would also have excellent sensitivity to an incident, relativistic,
highly-ionizing
magnetic monopole. Briefly, at large distances from a radio array, the
monopole will leave a trail of ionization, boosted in magnitude relative to
a muon by the large monopole charge. This trail of ionization is less
susceptible to LPM effects than a single UHE neutrino since the monopole
energy loss is distributed over a long pathlength. As shown in Figure
\ref{fig:monopole}, the preliminary limits from RICE are competitive.

Perhaps the primary virtue of the
in-ice strategy is the ability to vertex sources within the ice
itself.
Over the next several years,
AURA will (hopefully) take advantage of the timely opportunity 
presented by the IceCube drilling to extend the RICE strategy.
The AURA array
will serve as a prototype for a future radio+acoustic
hybrid array, over a surface area of 100-$km^2$ at the South Pole,
centered around IceCube. The AURA strategy is to re-package the 
IceCube digital optical modules with radio electronics, with four
breakouts to a cluster of antennas. 

\subsubsection{Radio detection in salt}
As reported by Amy Connolly at this conference,
SALSA will employ the {\it in situ} approach, using salt as the 
target medium rather than ice. The site being studied most closely is
Cote Blanche Dubois in Mississippi, where signals have been
observed propagating through
300 meters of dome salt at a depth of approximately 500 m. 
Advantages of using salt as a target are its higher
density compared to ice (i.e., greater likelihood of contained
events) its year-round accessibility and the likelihood that water
and metal in the soil layer above the dome provides RF insulation;
disadvantages are the limited, and often irregular volume comprised
by salt domes, high drilling costs compared to ice (\$1M/hole compared to
\$50K for a 1km deep, 15 cm diameter hole in ice). Nevertheless,
if the attenuation length can be shown to be of order 500 meters
in the 100 MHz - 1 GHz interval, salt may be an extremely promising
candidate, the large cost of a 15 kilochannel array ($\sim$250 M)
notwithstanding.

\subsubsection{Planned Surface Arrays}
ARIANNA\cite{ARIANNA} is unique not only in its 
coupling radio antennas to the surface of the 
ice, but in its use of shelf ice, rather than interior compressed
snow as the target. This approach offers
the advantage of a non-varying index-of-refraction in
the target medium. In this scenario, the Ross Ice Shelf will be
populated with a large array of down-looking horn antennas; the
limited depth of the shelf ice ($\sim$250 m) is compensated for by
the expectation that signals resulting from distant interactions will
multiply reflect within the shelf ice 'waveguide' before being
captured in the surface horn antennas. Signal communications will be
patterned after the wireless AUGER model. Measurements in December
2006 will quantify the attenuation length of the shelf ice at 
radio frequencies using techniques similar to those used to measure
the $\sim$1.7 km radiofrequency attenuation length of ice at the 
South Pole.

An elevated array, at a height $h$ would view the horizon out to
a distance $l=\sqrt{2r_Eh}$; with $r_E=6360$ km, an array of
50-meter high, wind-powered 
towers at Vostok would view the cold, 3.2 km thick ice
there out to a distance of 25 km. Provided the problem of 
finding bearing lubricants functional in the --80 C winter
temperatures can be solved, RICE could be recommissioned as
ROAST (``RICE On A STick'') in 2-3 years.

\subsubsection{Hardware}
\begin{figure}[h]
\begin{minipage}{18pc}
\includegraphics[width=14pc,angle=-90]{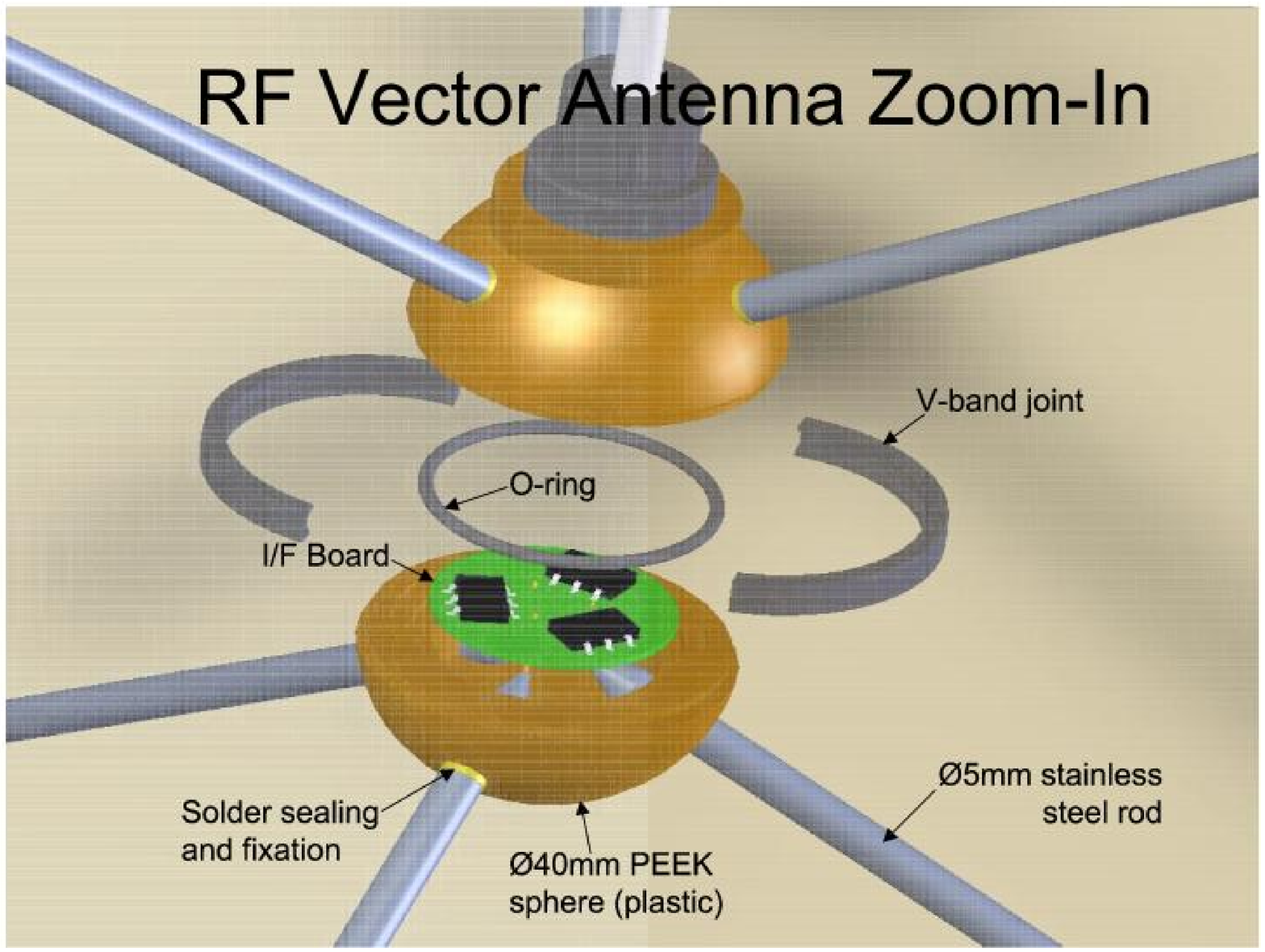}
\caption{\label{fig:tripol}Proposed AURA tripolarization antenna.}
\end{minipage}
\hspace{1pc}
\begin{minipage}{18pc}
\includegraphics[width=16pc,angle=0]{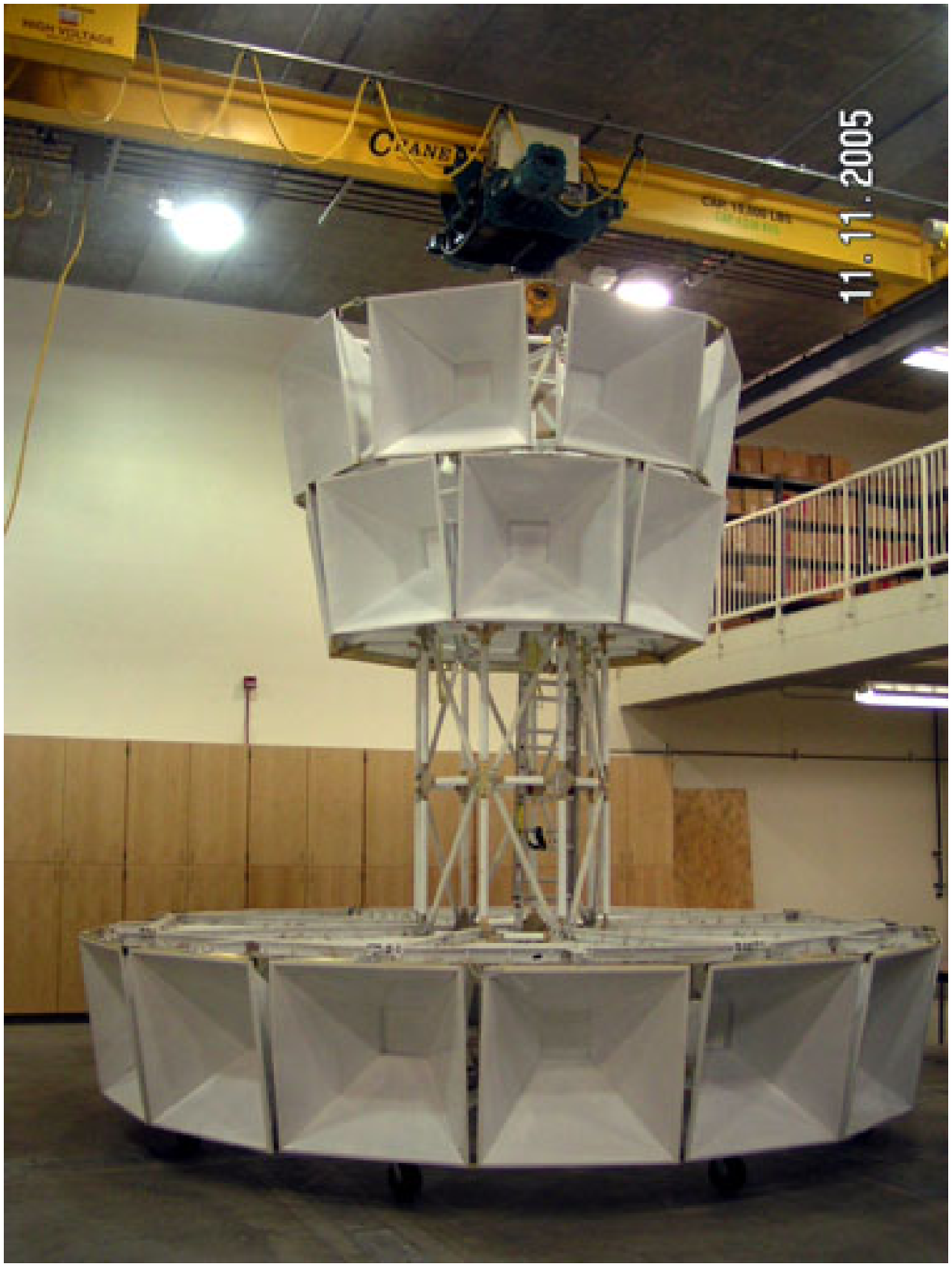}
\caption{\label{fig:dualpol}ANITA with complete suite of 36 dual-polarization horns.}\end{minipage} 
\end{figure}

Crucial to the success of any radio experiment is a highly efficient
antenna and a fast (ns-time scale), 
high-bandwidth data acquisition system\cite{labrador-ref}.
Both the tripolarization scheme planned for AURA (Fig. \ref{fig:tripol}), 
as well as the
dual-polarization horns that will fly on the ANITA gondola
(Fig. \ref{fig:dualpol}) will
offer excellent Cherenkov geometry constraints

ANITA have developed custom digitizing electronics, based
on a switched-capacitor array; triggering is `launched' by
an external trigger. This system has already demonstrated
1.3 GHz bandwidth at 2 GSa/sec, and will form the basis of the
future RICE successor array (AURA), as well. Frequency banding of the input
trigger signal allows rejection of possible
continuous wave backgrounds, as well as consistency with
the expected Askaryan frequency spectrum.
A multi-tiered trigger also facilitates
rejection of random thermal noise hits (by requiring at least
a two-fold coincidence at L1) or, in the case of a
buried array, rejection of down-coming anthropogenic transients.

\section{Radio detection of air showers} 
Coherent radio Cherenkov studies go back to Allan and
Jelly, who first studied radio signals from
cosmic ray air showers\cite{Allan,Jelly}.  
As an air shower develops in the earth's magnetic field, a lateral
separation of electric charge, of typical scale 10 meters (f$\sim$30 MHz) 
is produced by the Lorentz force, resulting
in a dipole field spiraling in towards the Earth for the orientation where the 
charges gyrate around ${\vec B}_{Earth}$, or simple charge separation when
the field lines are parallel to the Earth's surface (horizontal) 
and the charges
are incident along the vertical. The field strength is
obviously
a strong function of the inclination angle of the shower relative to the
earth's magnetic field -- both the azimuthal ($\phi$) and 
polar ($\theta$) angles, as well as the orientation and strength of the
local geomagnetic field are all important. The shower detection threshold is
set by various factors. Background-limiting factors include 
proximity to anthropogenic noise source
in the tens of Megahertz frequency regime. The ambient galactic radio noise is
also considerably larger than kTB thermal noise in this regime.
It is additionally noteworthy
that solar flares, auroral storms and lightning have all been observed to
result in substantial transient bursts in this frequency regime\cite{LOPES}. 
Signal-limiting factors include the typical distance from shower max to the
ground based observatory ($\sim$500 $g/cm^2$, or of order 5 km).

\subsection{Experimental Air Shower Detection Efforts}
After an extended hiatus, there is
a new generation of air-shower experiments which are now coming online. 
To date, anthropogenic backgrounds have limited the ability of such
arrays (thus far, modest in size) to self-trigger, generally requiring a
coincidence trigger with a co-located particle detector array to keep
such backgrounds manageable. As realized since the
1940's, the photomultiplier tubes
often used to readout scintillator arrays are,
unfortunately, themselves often a source
of considerable noise, which is not surprising given the few
ns rise-time of typical photomultiplier tube signals.

\begin{figure}[h]
\begin{minipage}{18pc}\includegraphics[width=16pc,angle=0]{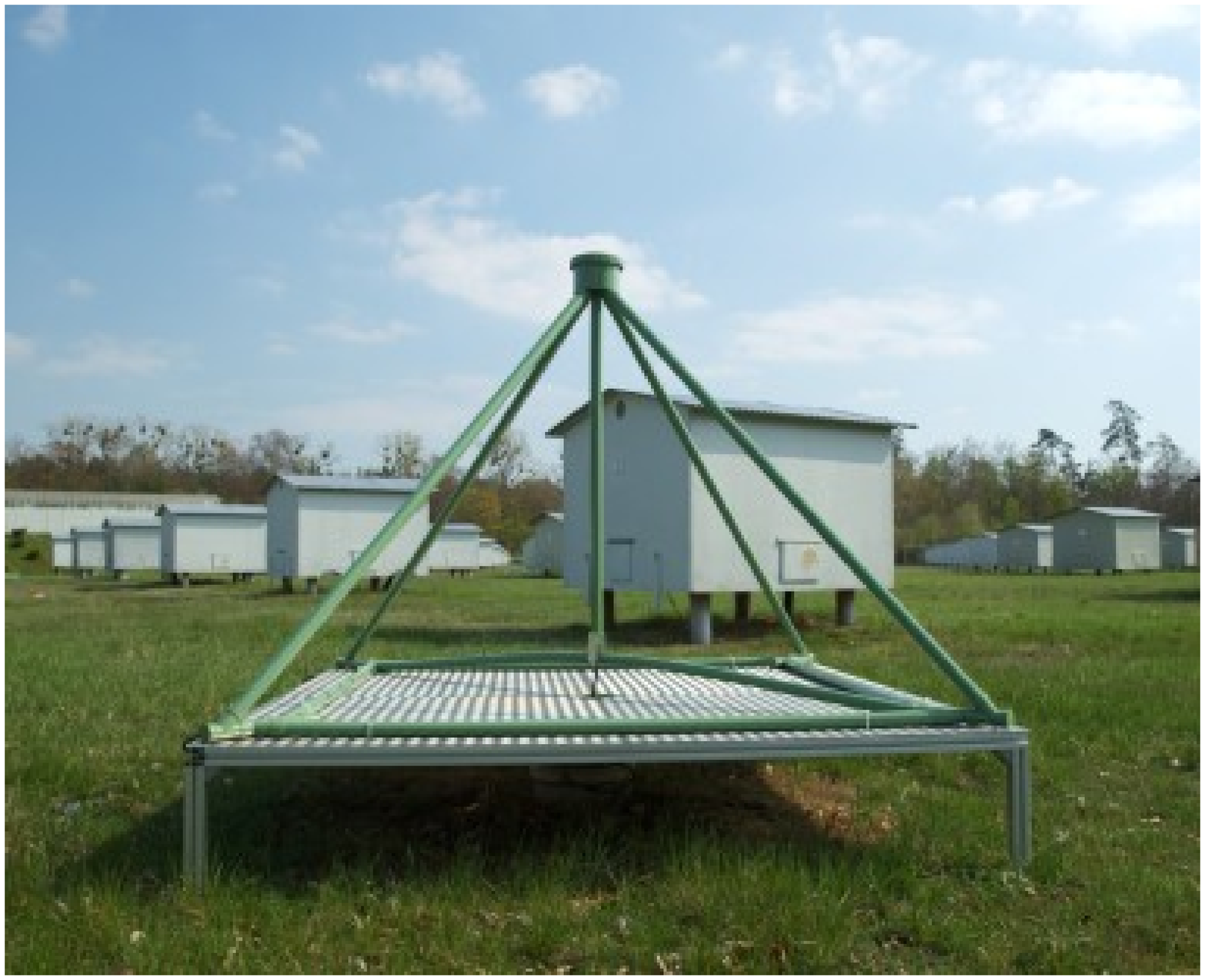}\caption{\label{fig:lopesant}Basic antenna element of LOPES array.}\end{minipage} 
\hspace{2pc}
\begin{minipage}{18pc}
\includegraphics[width=16pc,angle=0]{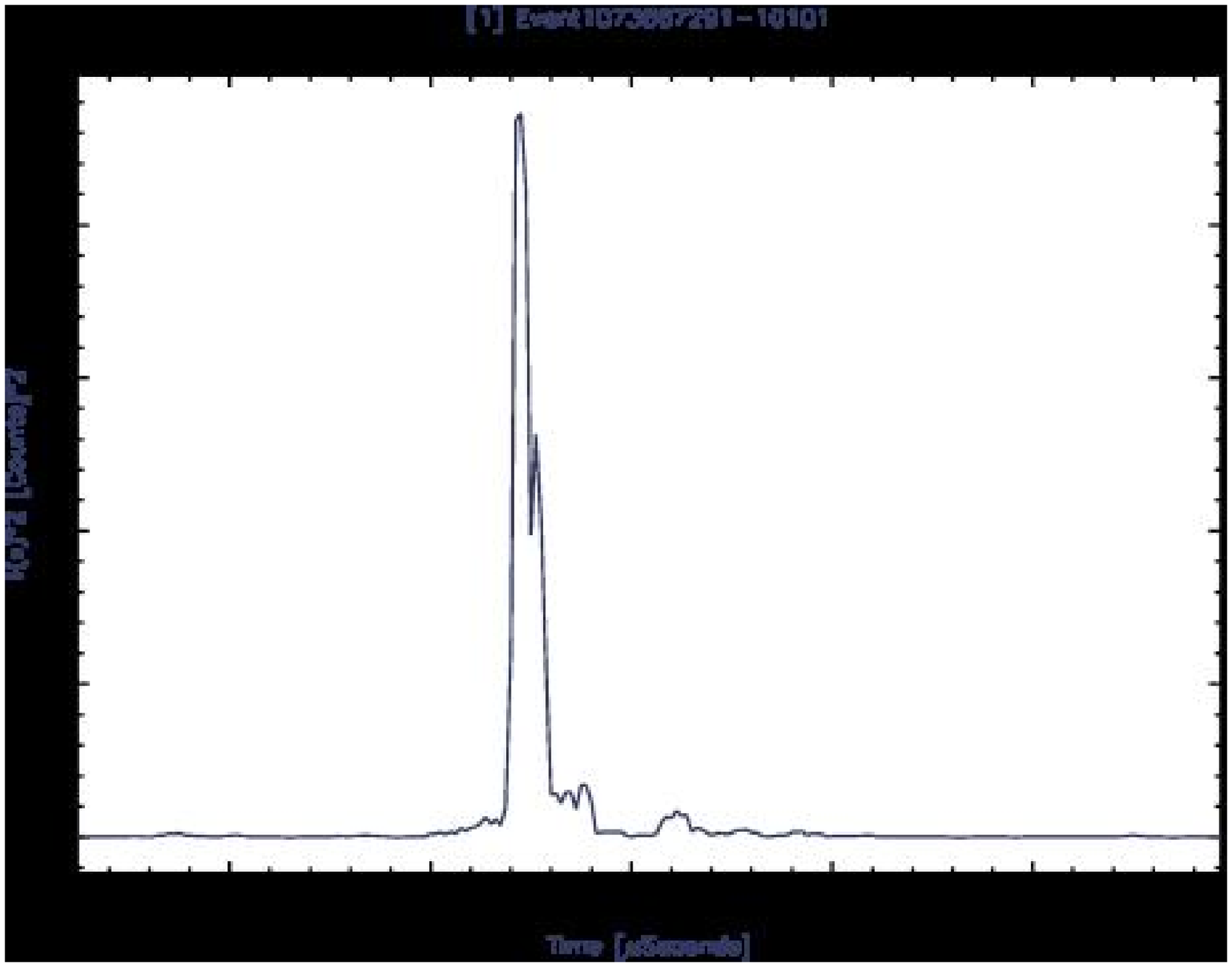}
\caption{\label{fig:signalpulselopes}Total radio power observed by LOPES, 2004 data.}
\end{minipage}
\end{figure}

Efforts within the last decade include those
of Rosner and Wilkerson at the Fly's
Eye/HiRes site (Dugway AFB) in Utah\cite{rosnerwilkerson}, 
the LOPES\cite{LOPES} prototype at the KASCADE site in Karlsruhe, Germany,
which is intended to serve as a prototype for an adaption of LOFAR
for air shower detection, and the CODALEMA\cite{CODALEMA} 
experiment in France.
There have also been discussions regarding the augmentation of
IceCube/ICETOP with a radio air shower array, and the
development of a similar array at the TUNKA site at Lake Baikal, Russia.
Of these, the LOPES and CODALEMA efforts are currently the most
advanced. The first LOPES data collection was performed in 2004 on a 
ten-element array consisting of linearly polarized, east-west oriented
``inverted V'' dipole
antennas, sensitive over the interval 40-80 MHz, 
embedded within (and triggered by) KASCADE. Seven months of data-taking, 
with an effective air shower threshold of 50 PeV,
yielded 862 candidate coincidence events, of which approximately 
40\% turn out to be well-reconstructed. Imposing the condition
of full radio coherence can substantially improve the angular
resolution and shower parameter estimates just based on the
ground array data; improvements of factors of 2-3 are not uncommon.
The basic element of the current LOPES-30 array and the 
observed signal are shown in Figures \ref{fig:lopesant}
and \ref{fig:signalpulselopes}, respectively. Ultimately,
one would like to include air shower detection trigger electronics
for the full, planned 25000 element LOFAR array. In the interim, a
radio air shower detector may be added to the
AUGER observatory.

The French CODALEMA collaboration uses 11 log-periodic antennas
(37-70 MHz) for
air shower detection, also with an energy threshold of 50 PeV.
Initially, a four-fold coincidence of
particle detectors on the ground were required to
provide an event trigger; since then, signal
recognition algorithms have advanced to allow for 
self-triggering by the radio array alone. CODALEMA currently
observes $\sim$1 high-energy air shower per day, and is planning
an upgrade to supplement the current log-periodic array with dipoles.

\section{Radio observations of astronomical objects}
As originally proposed by Askaryan, the moon provides a convenient,
otherwise radio-quiet,
and nearby,
``target-in-the-sky''. Neutrino collisions with moon rock would
result in a coherent radio pulse escaping through the surface.
In the absence of an atmosphere, ultra-high energy cosmic rays
would result in a similar signal. The Puschino observatory off the
banks of the Osa River, approximately 100 km south of Moscow, has
had a long history of lunar observation. More recently, the Goldstone
dish was adapted with trigger and data-acquisition electronics
appropriate for measurement of a neutrino-generated radio signal\cite{GLUE}.
At this conference, the nuMoon\cite{nuMoon} project was described, using as
receivers the Westerbork antenna array in Europe, consisting of 
14 25-m diameter radio dishes, with reception peaking at $\sim$150 MHz.
Given the very precise angular resolution anticipated for LOFAR,
Westerbork could serve as a testbed for the signal recognition
algorithms that would be used for such an extensive radio array.
All these experiments typically use the ability to frequency band
to discriminate the electric field spectral properties expected 
for Askaryan-type
signals ($dE/d\omega\sim\omega$) from backgrounds. Note that,
despite the fact that the electric field strength is increasing
with frequency, the higher angular spread of low-frequency components
of the Askaryan pulse result in more detections with an array
tuned to 150 MHz rather than 1.5 GHz.

\section{Summary of Extant and Planned Experiments}
The various techniques are summarized in the table below, showing
approximate thresholds, and neutrino flux sensitivity in a given
amount of livetime. %Projected sensitivities for a selection of experiments is given in Figure \ref{fig:ariannasens}.
\begin{table}[htpb]
\begin{tabular}{c|c|c|c|c} \\ 
Experiment & Status & Antennas & $E_\nu^0$ (PeV) & Comment \\ \hline
ANITA & 2006 Launch & 36 dual-pol horn 
& $10^4$ & 45-day 06-07 flight \\ \hline
ARIANNA & R\&D & 10000 horn & 100 & $L_{atten}$ TBD Dec., 06 \\ \hline
AURA & R\&D & (36 cluster)(4 Rx/cluster) & 100 & RICE successor\\ \hline
RICE & active & 20 in-ice dipole & 100 & data-taking$\to$2008 \\ \hline
SALSA & R\&D & 14000 dipole & 100 & Cote Blanche, MS \\ \hline
CODALEMA & active & 11 dipole & 100 & air showers \\ \hline
LOPES & active & 30 dipole & 100 & air showers \\ \hline
nuMoon & planned &  14 25-m dishes & $10^5$ & \\ \hline
\end{tabular}
\end{table}

\section{Backgrounds}
\begin{figure}[h]
\begin{minipage}{16pc}
\includegraphics[width=12pc,angle=0]{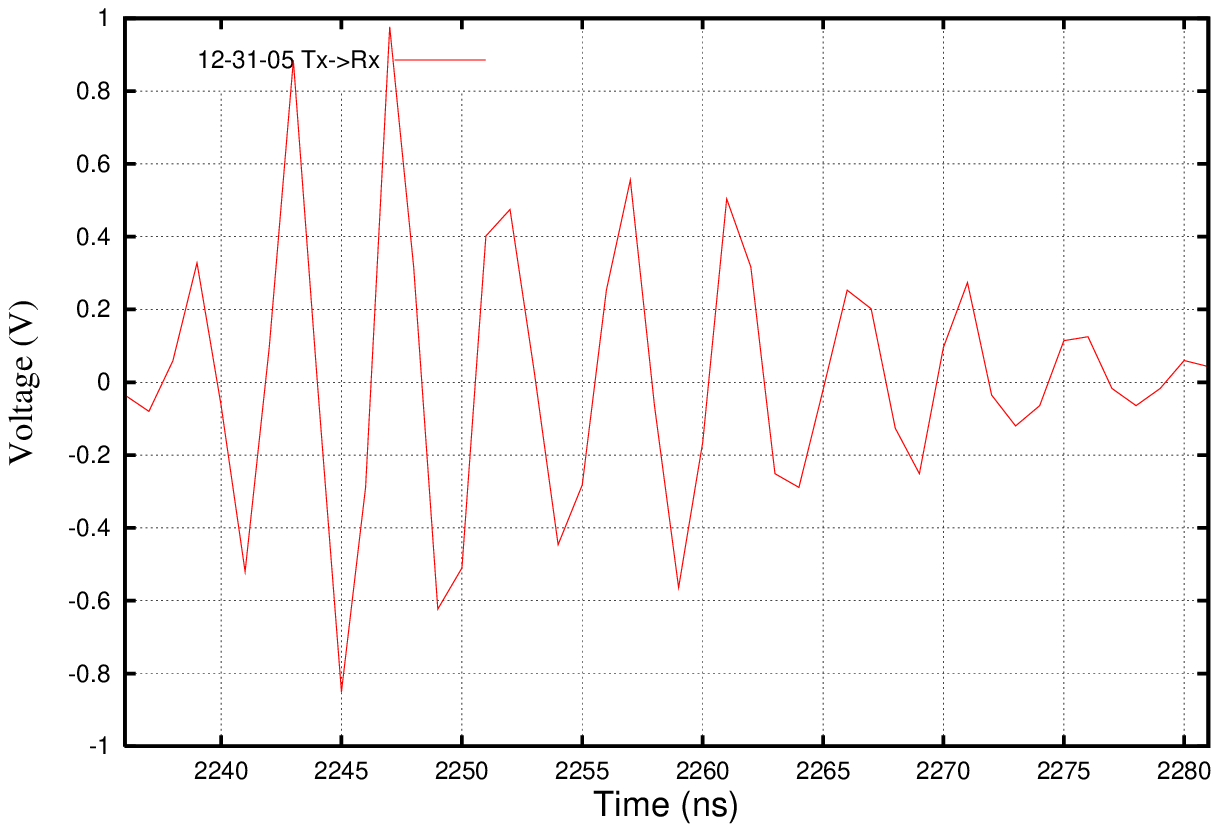}
\includegraphics[width=12pc,angle=0]{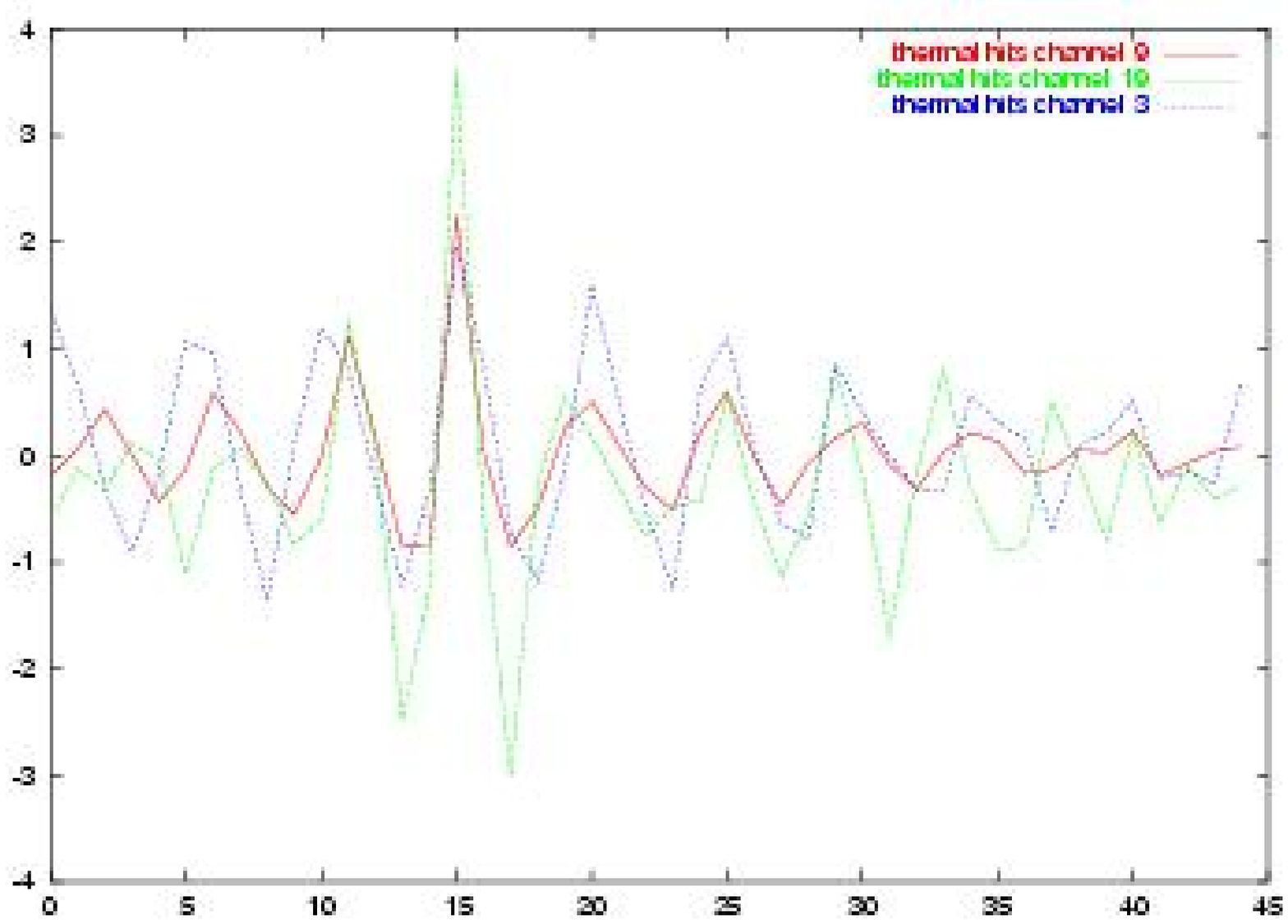}\caption{\label{fig:thermal}Top: Captured waveform for in-ice RICE Tx 
broadcasting to in-ice receiver. Bottom: putative RICE
``thermal'' noise hit.}
\end{minipage}
\hspace{0pc}
\begin{minipage}{18pc}
\includegraphics[width=18pc,angle=0]{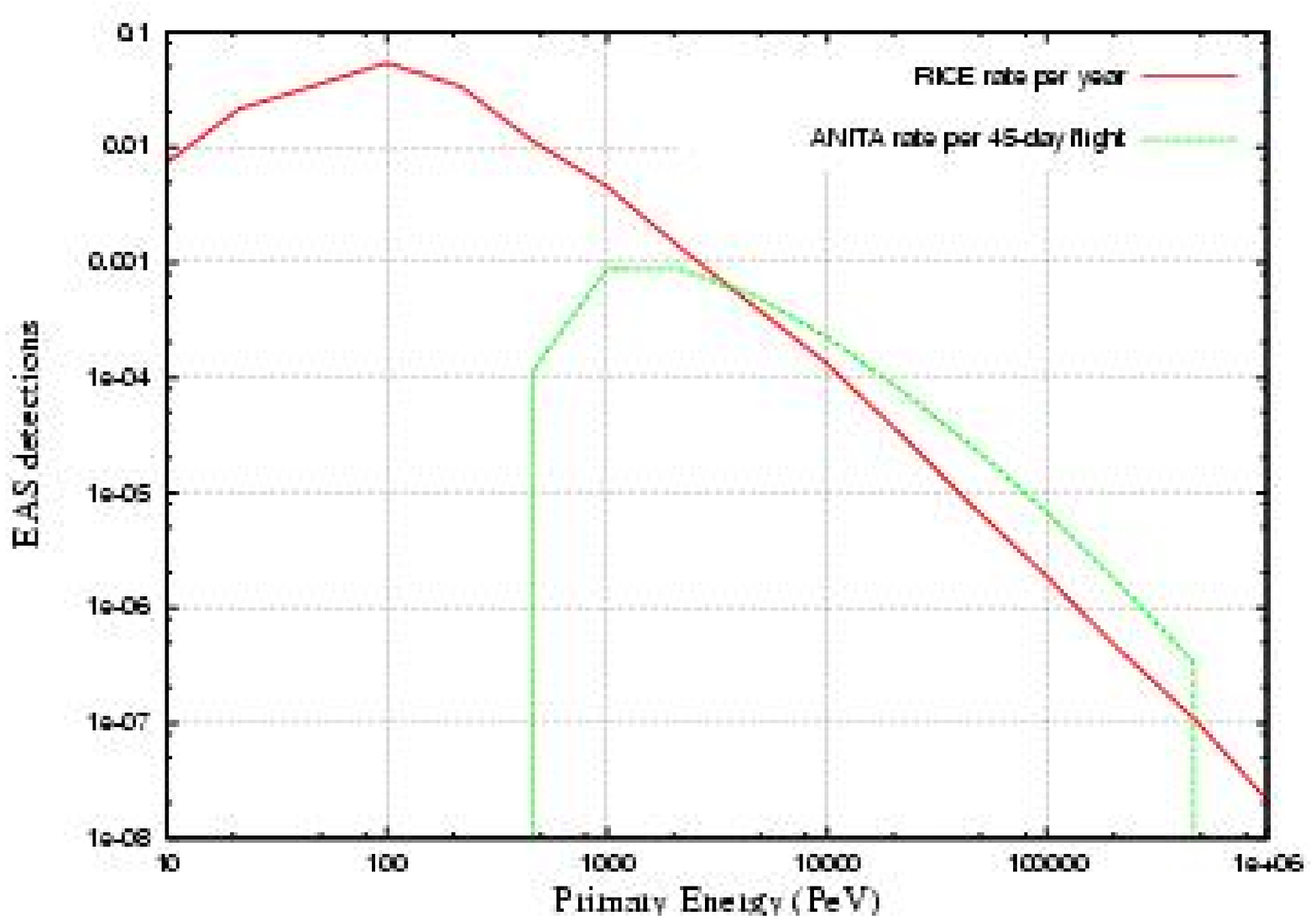}
\caption{\label{fig:EAS}Estimated EAS detections per year for RICE
and ANITA.}
\end{minipage}
\end{figure}

Perhaps the main advantage of the optical technique is that the 
primary background (atmospheric neutrinos) is, with the exception
of angular distribution, otherwise identical to the signal. In the
absence of atmospheric neutrinos at $>$PeV energies,
radio backgrounds are, in general, locale-specific. At the 
South Pole, in proximity to South Pole Station, the RICE backgrounds
are anthropogenic and can quickly ($\sim$1 microsecond) identified
as originating from the surface with $>$99\% efficiency. 
Backgrounds arising from thermal noise, compounded by the
system temperature of active electronics, present the most obvious
single-channel `fake' signals, and define the minimum voltage
threshold required to run the data acquisition system at tolerable
rates; at first glance, such backgrounds can look very similar
to `true' signals (Fig. \ref{fig:thermal}). 
Galactic radiofrequency noise has a falling power
spectrum and can be explicitly
suppressed by use of a 200 MHz high-pass filter.

Physics backgrounds to the
ice-based experiments at hundreds of PeV are $expected$ to be small. 
Many processes have been considered -- direct radiofrequency
signals from extensive air showers (EAS)
which propagate into the ice, as measured by the CODALEMA
and LOPES experiments, give signals which
peak in the tens of MHz regime. There should be a signal
resulting from the impact of the shower core with the ice,
producing the same kind of Cherenkov radiation signal that
RICE seeks to measure. Nevertheless (Fig \ref{fig:EAS}), the
expected EAS signal rate should be almost immeasurably small.

\section{Comparison with acoustic}
Acoustic detection of neutrinos has undergone equally rapid
development over the last few years, as outlined
elsewhere\cite{SAUND,rolfreview,sebastianthesis}. 
The acoustic detection technique offers three very large advantages
relative to radio detection: a) propagation speeds, and absorption
lengths are set by the acoustic properties of the target medium;
for ice, the slow propagation speeds mean that signals are spread over
timescales of microseconds and are therefore $1000\times$ less
exacting than the nanosecond time scales of radio pulses. Additionally,
at these low frequencies, cable losses are sufficiently small that
signals can be digitized in an accessible 
surface, rather than an inaccessible in-ice receiver. 
b) the signal geometry is in a domain where the source length is
much larger than the wavelength of the propagating signal, so that
the signal geometry is cylindrical, with amplitude losses as 
$\propto 1/ln(r)$, and c) for ice, the attenuation length is expected
to be a factor $10\times$ larger than radio; for salt, estimates give
values an additional factor of $10\times$ larger\cite{bufordsalt}. 
Both techniques
offer the possibility of constraining the signal geometry using
polarization information, with the exception of acoustic detection
in salt, for which coupling to the shear wave will be difficult. The long
time scale for the signal leads,
however, to the one large disadvantage of acoustic - the substantially larger
1/f noise for acoustic detection forces the detection threshold up to
around 100 EeV or so.

\section{Perspectives}
Although promising, the radio technique is still being developed, and still 
must surmount several obstacles before gaining widespread acceptance. 
The problem is not so much signal detection, but understanding backgrounds.
The air detection experiments have made enormous strides within the last
five years -- the demonstrated ability to measure 
the same air shower event as a ground array
provides instant credibility, however, anthropogenic backgrounds and
naturally generated noise such as electrical discharges in the atmosphere
require frequency filtering, as well as restrictive data selection. The
next milestone here may be an absolutely normalized air shower
primary energy spectrum which
shows good agreement with the existing data. The demonstrated ability to
self-trigger is also a {\it condicio sine qua non} for a self-sustaining array.
In the absence of such self-triggering, the ability to improve the angular
resolution of air showers is, nevertheless, sufficient cause for expansion
of this effort. Estimates of
neutrino detection in dense media 
rests on three primary pillars: a) the radio transparency of the medium,
b) the weak interaction cross-section at high energies, and c) 
the relationship between shower characteristics and radio signal strength.
The attenuation length of Antarctic ice has been measured to be $\sim$1.5-2 km,
$\sigma_\nu$ can be extrapolated 
within the Standard Model to within 25\% or so at 10 EeV, and
the SLAC
testbeam experiments have (I believe, conclusively) validated the
signal strength estimates obtained using Monte Carlo simulations.
Thermal backgrounds can, in principle, be removed
statistically based on channel-to-channel correlations
and also frequency spectra, however, our experience with RICE indicates
that individual-channel thermal 'hits' can be extremely similar to 
expected neutrino 'signal' hits, as shown in Figure \ref{fig:thermal};
large bandwidth is required to separate the two
in the frequency domain.

Clearly, a single experiment with the
potential of simultaneously offering
detection of air showers, both with radio and a conventional
ground particle detector array, plus
detection of showers-in-dense-media, combining optical + acoustic + radio
capabilities, would be perhaps the most powerful realization of
a 'multi-messenger' detector. 
Obviously, the optimal scenario would be a single laboratory which could
support simultaneous measurement of optical + radio + acoustic signals,
with all possible polarizations. Currently, several groups are working
towards the realization of such an observatory at the South Pole 
(``CONDOR'') which,
given, e.g., the inability to propagate radio waves through water,
is likely the only place on Earth capable of supporting such an effort.
Mature simulations indicate measurable registered neutrino coincidences
in one year\cite{justin}.

\section{Acknowledgments}
I would like to thank everyone who is currently active in this very dynamic
field. I am particularly indebted to Peter Gorham, Daniel Hogan, 
Shahid Hussain, Soeb Razzaque,
and David Saltzberg for 
their contributions to this talk.

\end{document}